\documentclass[12pt,a4page]{amsart} 
\usepackage{amsmath,amssymb,amsfonts,amsthm,array}
\usepackage{listings}
\usepackage{color}
\usepackage{amsaddr}
\usepackage{hyperref}
\usepackage{graphicx}

\definecolor{dkgreen}{rgb}{0,0.6,0}
\definecolor{gray}{rgb}{0.5,0.5,0.5}
\definecolor{mauve}{rgb}{0.58,0,0.82}
\calclayout

\def\EE{{\mathbb E}}

\def\XXX{{\mathbf X}}
\def\ZZZ{{\mathbf Z}}

\def\PPP{{\mathbf P}}

\def\RR{{\mathbb R}}
\def\rr{\mathbb R}

\title{\MakeLowercase{rcss} : Subgradient and duality approach for
  dynamic programming}

\author{Juri Hinz \and Jeremy Yee*} \email{jeremyyee@outlook.com.au}

\begin{document}

\maketitle

\begin{abstract} 
  This short paper gives an introduction to the \emph{rcss} package.
  The R package \emph{rcss} provides users with a tool to approximate
  the value functions in the Bellman recursion using convex piecewise
  linear functions formed using operations on tangents. A pathwise
  method is then used to gauge the quality of the numerical results.
\end{abstract}

\smallskip
\noindent \textbf{Keywords.}  Convexity, Dynamic programming, Duality,
Subgradient

\section{Introduction}

Sequential decision making is often addressed under the framework of
{\it Markov Decision Processes/Dynamic Programming}. However, deriving
analytical solutions for even some of the simplest decision processes
may be too cumbersome \cite{powell, bauerle_rieder,pham}. The use of
numerical approximations may be far more practical given the rapid
improvements in everyday computational power. The ability to gauge the
quality of these approximations is also of significant practical
importance. This paper will describe the implementation of fast and
accurate algorithms to address these issues for Markov decision
processes within a finite time setting, finite action set, convex
reward functions and whose Markov processes follow linear
dynamics. Under certain conditions, \cite{hinz2014} showed that these
value function approximations enjoy uniform convergence on compact
sets. The package \emph{rcss} represents a \emph{R} implementation of
these methods and has already been used to address problems such as
pricing financial options \cite{hinz_yap2015}, natural resource
extraction \cite{hinz_etal_mining}, battery management
\cite{hinz_yee_battery}, optimal portfolio liquidation
\cite{hinz_yee_liquidation} and optimal asset allocation under hidden
state dynamics \cite{hinz_yee_pomdp}. One of the major benefits of
implementing these methods in \emph{R} \cite{r} is that the results
can be analysed using the vast number of statistical tools avaliable
in this language.  The \emph{R} package can be found here:
\url{https://github.com/YeeJeremy/rcss} and the manual is listed at
\url{https://github.com/YeeJeremy/RPackageManuals/blob/master/rcss-manual.pdf}.

\section{Problem Setting}
\label{sec_problem}

Suppose that state space $\XXX = \mathbf{P} \times \ZZZ$ is the
product of a finite set $\mathbf{P}$ and an open convex set
$\ZZZ\subseteq \mathbb{R}^d$.  At each decision time
$t \in \{0, 1, \dots, T-1\}$, an action $a \in \mathbf{A}$ is chosen
by the agent and the dynamic choice of these actions influences the
evolution of the Markov process
$(X_t)_{t=0}^T := (P_t,Z_t)_{t=0}^T: \Omega \rightarrow \mathbf{P}
\times \ZZZ$ where $\Omega$ is the set of sample paths. The discrete
component $(P_t)_{t=0}^T$ is assumed to be a controlled Markov chain
with transition probabilities
$(\alpha_{p,p'}^a)_{p, p' \in \mathbf{P}}, a \in \mathbf{A}$, where
$\alpha_{p,p'}^a$ is the probability of transitioning from $p$ to $p'$
after applying action $a$. The second component $(Z_t)_{t=0}^T$
evolves in a linear fashion given by $Z_{t+1} = W_{t+1} Z_t$ where
$(W_t)_{t=1}^{T}$ are matrix-valued random variables refered to as
disturbances. The matrix entries in these disturbances are assumed to
be integrable. At each time $t=0, \dots, T-1$ the decision rule
$\pi_{t}$ is given by a mapping $\pi_{t}: \XXX \to \mathbf{A}$,
prescribing at time $t$ an action $\pi_{t}(p,z) \in \mathbf{A}$ for a
given state $(p, z) \in \XXX$. A sequence
$\pi = (\pi_{t})_{t=0}^{T-1}$ of decision rules is called a policy.
For each policy $\pi = (\pi_{t})_{t=0}^{T-1}$, associate it with a
so-called policy value $v^{\pi}_{0}(p_0, z_{0})$ defined as the total
expected reward
\begin{equation*}
v^{\pi}_{0}(p_0, z_{0})=\mathbb{E}^{x_0,\pi}\left[\sum_{t=0}^{T-1}
r_{t}(P_t, Z^{}_{t}, \pi_{t}(X_t)) + r_T(P_T,Z_T)\right]
\end{equation*}
where $r_T:\PPP\times\ZZZ\to\RR$ and
$r_t: \mathbf{P} \times \ZZZ \times \mathbf{A} \rightarrow \mathbb{R}$
are convex functions in the second argument for $t=0,\dots,T-1$. These
functions represent the scrap and reward in the decision problem,
respectively. A policy $\pi^{*}=(\pi^{*}_{t})_{t=0}^{T-1}$ is called
optimal if it maximizes the total expected reward over all policies
$\pi \mapsto v^{\pi}_{0}(p, z)$. To obtain such policy, one introduces
for $t = 0, \dots, T-1$ the so-called {\it Bellman operator}
\begin{equation*} 
{\mathcal T}_{t}v(p,z)=\max_{a \in \mathbf{A}} \left\{r_{t}(p,z ,
a) + \sum_{p'\in\mathbf{P}} \alpha_{p,p'}^a\mathbb{E}^W[v(p',W_{t+1}z)]\right\}, 
\quad (p, z) \in \mathbf{P} \times \ZZZ \label{bell}
\end{equation*} 
acting on all functions $v$ where the expectation is defined. Consider
the {Bellman recursion}, also referred to as backward induction:
\begin{equation*} 
  v_{T}(p, z)= r_{T}(p,z), \quad v_{t}=
  {\mathcal T}_{t} v_{t+1} \qquad \hbox{for $t=T-1, \dots,
    0$.} \label{backward_induction}
\end{equation*}
A recursive solution $(v^{*}_{t})_{t=0}^{T}$ to the Bellman recursion
above are called {value functions} and they determine an optimal
policy $\pi^{*}=(\pi^{*}_{t})_{t=0}^{T-1}$ via
\begin{equation*}
\pi^{*}_{t}(p,z) = \arg\max_{a \in \mathbf{A}}\left\{r_{t}(p, z, a)+
\sum_{p'\in\mathbf{P}}
\alpha_{p,p'}^a\mathbb{E}^W[v^{*}_{t+1}(p',W_{t+1}z)] \right\},
\end{equation*}
for $t = T-1,\dots, 0$.

\section{Numerical Approach}
\label{sec_algo}

Since the reward and scrap functions are convex in the continuous
variable, the value functions are also convex due to the linear state
dynamics and so can be approximated by convex piecewise linear
functions. For this, introduce the so-called subgradient envelope
${\mathcal S}_{\mathbf{G}^m}f$ of a convex function $f: \ZZZ \to \rr$
on a grid $\mathbf{G}^m \subset \ZZZ$ with $m$ points
i.e. $\mathbf{G}^{m}=\{g^{1}, \dots, g^{m}\}$ by
$$
{\mathcal S}_{\mathbf{G}^m}f=\vee_{g \in \mathbf{G}^m}
(\triangledown_{g}f)
$$
which is a maximum of the tangents $\triangledown_{g}f$ of $f$ on all
grid points $g \in \mathbf{G}^m$. Using the subgradient envelope
operator, define the double-modified Bellman operator as
$$
{\mathcal T}^{m, n}_{t}v(p, \cdot) = {{\mathcal S}_{\mathbf{G}^{m}}} {\max_{a \in
\mathbf{A}}}\left( {r_{t}(p, \cdot , a)+}
{\sum_{p'\in\mathbf{P}} \alpha_{p,p'}^a\sum_{k=1}^{n}\nu^{(k)}_{t+1}v(p', W^{(k)}_{t+1}\cdot)} \right)
$$
where the probability weights $(\nu^{(k)}_{t+1})_{k=1}^{n}$ corresponds
to the distribution sampling $(W_{t+1}^{(k)})_{k=1}^{n}$ of each
disturbance $W_{t+1}$. The corresponding backward induction
\begin{eqnarray*} 
  v^{m, n}_{T-1}(p, z)&=& \mathcal{T}^{m,n}_{T-1}{\mathcal S}_{\mathbf{G}^{m}} r_{T}(p, z), \label{scheme1}\\
  v^{m, n}_{t}(p, z)&=&{\mathcal T}^{m, n}_{t}v^{m, n}_{t+1}(p, z),
                        \qquad t=T-2, \dots 0. \label{scheme2}
\end{eqnarray*}
for $p\in\PPP$ and $z\in\ZZZ$ yields the so-called double-modified
value functions $(v^{m, n}_{t})_{t=0}^{T}$. If the disturbance
sampling is constructed using local averages on a partition of the
disturbance space or using random Monte Carlo sampling, it can be
shown that the double-modified value functions converge uniformly to
the true value functions on compact sets if the grid becomes dense in
$\ZZZ$. Now, to gauge the quality of the approximations from the
above, we construct two random variables whose expectaions bound the
true value function i.e.
\begin{equation}
  \EE(\underline{\upsilon}_0(p,z_0)) \leq v_0(p,z_0) \leq \EE(\overline{\upsilon}_0(p,z_0)), \qquad p \in \mathbf{P}, \quad z_0 \in \ZZZ.
  \label{gap}
\end{equation}
This process exhibits a helpful {self-tuning} property. The the closer
the value function approximations to optimality, the tighter the
bounds in Equation \ref{gap} and the lower the standard errors of the
bound estimates. 

The R package \emph{rcss} represents these convex piecewise linear
functions as matrices and offers several options to use nearest
neighbour algorithms (from \cite{flann}) to reduce the computational
cost of the above methods.  Most of the computational work is done in
C++ via \emph{Rcpp} \cite{rcpp} and is parallezied using \emph{OpenMp}
\cite{openmp}. The following sections will demonstrate some real world
applications of this R package.

\section{Example: Bermuda Put}

Optimal switching problems naturally arise in the valuation of
financial contracts. A simple example is given by the Bermudan Put
option. This option gives its owner the right but not an obligation to
choose a time to exercise the option in order to receive a payment
which depends on the price of the underlying asset at the exercise
time. The so-called fair price of the Bermudan option is related to
the solution of an optimal stopping problem (see \cite{glasserman}).
Here, the asset price process $( \tilde Z_{t})_{t=0}^{T}$ at time
steps $0, \dots, T$ is modelled as a sampled geometric Brownian motion
\begin{equation*}
\tilde Z_{t+1}=  \epsilon_{t+1} \tilde Z_{t}, \quad t=0, \dots, T-1,  \,  Z_{0} \in \rr_{+}, \label{prev}
\end{equation*}
where $( \varepsilon_{t})_{t=1}^{T}$ are independent random variables
following a log-normal distribution.  The fair price of such option
with strike price $K$, interest rate $\rho\ge0$ and maturity date $T$,
is given by the solution to the optimal stopping problem
$$ 
\begin{array}{c}
\sup\{ \EE( \max(e^{-\rho \tau}(K- \tilde Z_{\tau}), 0) ): \enspace  
\hbox{$\tau$ is $\{0,1, \ldots, T\}$-valued stopping time}\}.
\end{array}
$$
A transformation of the state space is required to be able
representing the reward functions in a convenient way for the
\emph{rcss} package to process, thus we introduce an augmentation with
1 via
$$
Z_{t}=\left[\begin{array}{c} 1 \\ \tilde Z_{t} \end{array}\right], \qquad  t=0, \dots, T.
$$
then it becomes possible to represent the evolution as the linear
state dynamics
$$
Z_{t+1}=W_{t+1}Z_{t}, \qquad t=0, \dots, T-1
$$
with independent and identically distributed matrix-valued random
variables $(W_{t})_{t=1}^{T}$ given by
\begin{equation*}
W_{t+1}= \left[ \begin{array}{cc} 1 & 0 \\  0 & \epsilon_{t+1}
\end{array}\right], \quad t=0,...,T-1. \label{bm}
\end{equation*}

This switching system is defined by two positions
$\mathbf{P}=\{1, 2\}$ and two actions $\mathbf{A}=\{1, 2\}$. Here, the
positions `exercised' and `not exercised' are represented by $p=1$,
$p=2$ respectively, and the actions `don't exercise' and `exercise'
are denoted by $a=1$ and $a=2$ respectively. With this interpretation,
the position change is given by deterministic transitions to specified
states
\begin{equation*}
\alpha_{p,p'}^a
=\left\{
 \begin{array}{cl}
  1 &  \text{if $p'=\alpha(p, a)$ } \\ 0 & \text{else } \label{deftwodim1}
 \end{array}
\right.
\end{equation*}
deterministically determined by the  target positions 
\begin{equation*} \label{twodim1}
(\alpha(p, a))_{p,a=1}^{2} \sim
\left[\begin{array}{cc}
\alpha(1,1) & \alpha(1,2)\\
\alpha(2,1) & \alpha(2,2)
\end{array}
\right]
=\left[\begin{array}{cc}
1 & 1\\
2 & 1
\end{array}
\right],
\end{equation*}
while the rewards at time $t=0, \dots, T$  and are defined as
\begin{eqnarray*}
r_{t}(p, (z^{(1)}, z^{(2)}), a) &=& e^{-\rho t}\max(K-  z^{(2)}, 0)(p-\alpha(p, a)),  \label{rewput}\\
r_{T}(p, (z^{(1)}, z^{(2)})) &=& e^{-\rho T}\max(K-  z^{(2)}, 0)(p-\alpha(p, 2)), \label{scrapput}
\end{eqnarray*}
for all $p\in \mathbf{P}$,  $a \in \mathbf{A}$, $z \in \rr_{+}$.

\subsection{Code Example}

As a demonstration, let us consider a Bermuda put option with strike
price 40 that expires in 1 year. The put option is exercisable at 51
evenly spaced time points in the year, which includes the start and
end of the year. The following code approximates the value functions
in the Bellman recursion.  On a Linux Ubuntu 16.04 with Intel i5-5300U
CPU @2.30GHz and 16GB of RAM, the following code takes around 0.2 cpu
second and around 0.05 real world seconds.

\begin{lstlisting}[caption = {Value function approximation}]
library(rcss)
rate <- 0.06 ## Interest rate
step <- 0.02 ## Time step between decision epochs
vol <- 0.2 ## Volatility of stock price process
n_dec <- 51 ## Number of decision epochs
strike <- 40 ## Strike price
control <- matrix(c(c(1, 1), c(2, 1)), nrow = 2, byrow = TRUE)  ## Control
grid <- as.matrix(cbind(rep(1, 301), seq(30, 60, length = 301)))  ## Grid
## Disturbance sampling
u <- (rate - 0.5 * vol^2) * step
sigma <- vol * sqrt(step)
condExpected <- function(a, b){
    aa <- (log(a) - (u + sigma^2)) / sigma
    bb <- (log(b) - (u + sigma^2)) / sigma
    return(exp(u + sigma^2 / 2) * (pnorm(bb) - pnorm(aa)))
}
weight <- rep(1 / 1000, 1000)
disturb <- array(0, dim = c(2, 2, 1000))
disturb[1,1,] <- 1
part <- qlnorm(seq(0, 1, length = 1000 + 1), u, sigma)
for (i in 1:1000) {
    disturb[2,2,i] <- condExpected(part[i], part[i+1]) / (plnorm(part[i+1], u, sigma) - plnorm(part[i], u, sigma))
}
## Subgradient representation of reward
in_money <- grid[,2] <= strike
reward <- array(0, dim = c(301, 2, 2, 2, n_dec - 1))       
reward[in_money,1,2,2,] <- strike
reward[in_money,2,2,2,] <- -1
for (tt in 1:n_dec - 1){
    reward[,,,,tt] <- exp(-rate * step * (tt - 1)) * reward[,,,,tt]
}
## Subgrad representation of scrap
scrap <- array(data = 0, dim = c(301, 2, 2))
scrap[in_money,1,2] <- strike
scrap[in_money,2,2] <- -1
scrap <- exp(-rate * step * (n_dec - 1)) * scrap
## Bellman 
r_index <- matrix(c(2, 2), ncol = 2)
bellman <- FastBellman(grid, reward, scrap, control, disturb, weight, r_index)
\end{lstlisting}

The matrix \texttt{grid} represents our choice of grid points where
each row represents a point. The 3-dimensional array \texttt{disturb}
represents our sampling of the disturbances where
\texttt{disturb[,,i]} gives the i-th sample. Here, we use local
averages on a $1000$ component partition of the disturbance space. The
5-dimensional array \texttt{reward} represents the subgradient
approximation with \texttt{reward[,,a,p,t]} representing
${\mathcal S}_{{\mathbf G}^m} r_{t}(p,.,a)$. The object
\texttt{bellman} is a list containing the approximations of the value
functions and expected value functions for all positions and decision
epochs. Please refer to the package manual for the format of the
inputs and outputs. To obtain the value function of the Bermuda put
option, simply run the plot command below.
 
\begin{lstlisting}[caption = {Option value function}]
plot(grid[,2], rowSums(bellman$value[,,2,1] * grid), type = "l", xlab = "Stock Price", ylab = "Option Value")
\end{lstlisting}
\vspace{-5mm}
\begin{figure}[h!]
  \centering
  \includegraphics[height=3.5in,width=0.9\textwidth]{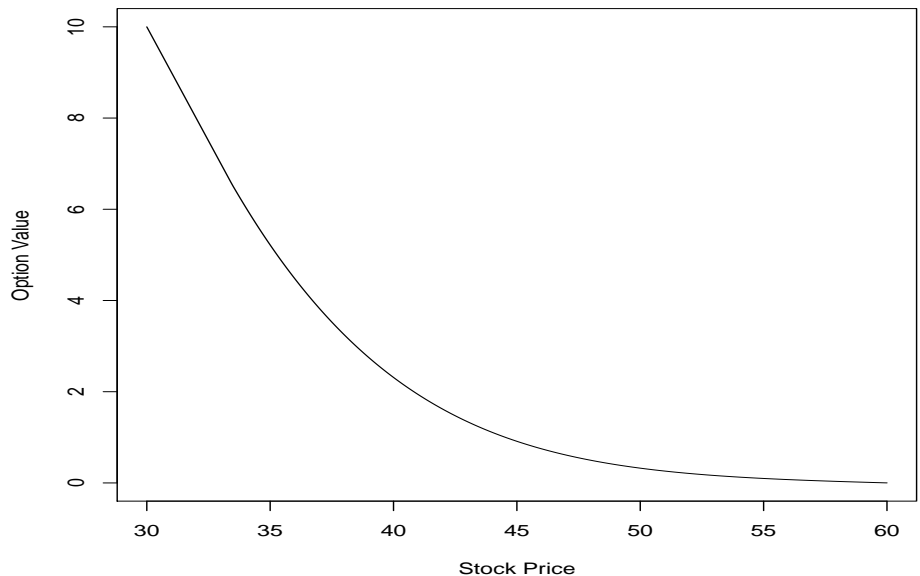}
  \caption{Bermuda put value function.}
  \label{bermuda_plot}
\end{figure}

The following code then computes the lower and upper bound estimates
for the value of the option when $\tilde Z_0 =
36$. On our machine, the following takes around
$10$ cpu seconds and around $5$ real world seconds to run.

\begin{lstlisting}[caption = {Lower and upper bounds}]
## Reward function
RewardFunc <- function(state, time) {
    output <- array(data = 0, dim = c(nrow(state), 2, 2))
    output[,2,2] <- exp(-rate * step * (time - 1)) * pmax(40 - state[,2], 0)
    return(output)
}
## Scrap function
ScrapFunc <- function(state) {
    output <- array(data = 0, dim = c(nrow(state), 2))
    output[,2] <- exp(-rate * step * (n_dec - 1)) * pmax(40 - state[,2], 0)
    return(output)
}
## Get primal-dual bounds
start <- c(1, 36)
## Path disturbances 
set.seed(12345)
n_path <- 500
path_disturb <- array(0, dim = c(2, 2, n_path, n_dec - 1))
path_disturb[1, 1,,] <- 1
rand1 <- rnorm(n_path * (n_dec - 1) / 2)
rand1 <- as.vector(rbind(rand1, -rand1))
path_disturb[2, 2,,] <- exp((rate - 0.5 * vol^2) * step + vol * sqrt(step) * rand1)
path <- PathDisturb(start, path_disturb)
policy <- FastPathPolicy(path, grid, control, RewardFunc, bellman$expected)
## Subsim disturbances
n_subsim <- 500
subsim <- array(0, dim = c(2, 2, n_subsim, n_path, (n_dec - 1)))
subsim[1,1,,,] <- 1
rand2 <- rnorm(n_subsim * n_path * (n_dec - 1) / 2)
rand2 <- as.vector(rbind(rand2, -rand2))
subsim[2,2,,,] <- exp((rate - 0.5 * vol^2) * step + vol * sqrt(step) * rand2)
subsim_weight <- rep(1 / n_subsim, n_subsim)
mart <- FastAddDual(path, subsim, subsim_weight, grid, bellman$value, ScrapFunc)
bounds <- AddDualBounds(path, control, RewardFunc, ScrapFunc, mart, policy)
\end{lstlisting}

The above code takes the exact reward and scrap functions as
inputs. The function \texttt{FastPathPolicy} computes the candidate
optimal policy.  The object \texttt{bounds} is a list containing the
primals $\underline{v}_t^i(p,z_t)$ and duals
$\underline{v}_t^i(p,z_t)$ for each sample path $i$ and each position $p$
at each decision time $t$. Again, please refer to the package manual
for the format of the inputs and outputs. If the price of the
underlying asset is $36$, the $99\%$ confidence interval for the
option price is given by the following.

\begin{lstlisting}[caption = {99\% confidence interval}]
> print(GetBounds(bounds, 0.01, 2))
[1] 4.475802 4.480533
\end{lstlisting}

The package \emph{'rcss'} also allows the user to test the prescribed
policy from the Bellman recursion on any supplied set of sample paths.
The resulting ouput can then be further studied with time series
analysis or other statistical work. In the following code, we will use
the previously generated $500$ sample paths to backtest our policy and
generate histograms.

\begin{lstlisting}[caption = {Backtesting Policy}]
test <- FullTestPolicy(2, path, control, RewardFunc, ScrapFunc, policy)
## Histogram of cumulated rewards
hist(test$value, xlab = "Cumulated Rewards", main = "")
## Exercise times
ex <- apply(test$position == 1, 1, function(x) min(which(x)))
ex[ex == Inf] <- 51
ex <- ex - 1
hist(ex, xlab = "Exercise Times", main = "")
\end{lstlisting}

\begin{figure}[h!]
  \centering
  \includegraphics[height=2.8in, width=0.45\textwidth]{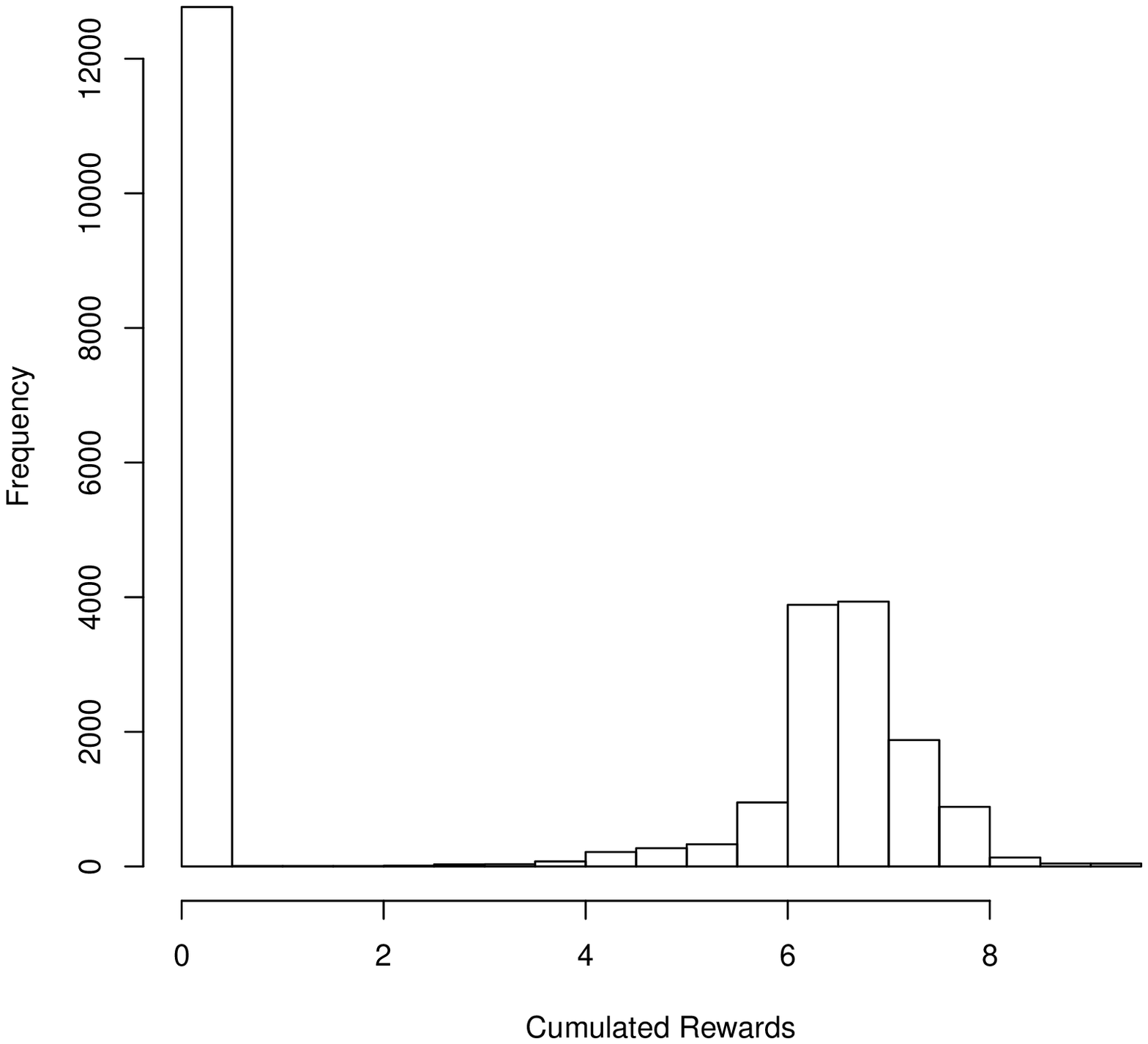}
  \includegraphics[height=2.8in, width=0.45\textwidth]{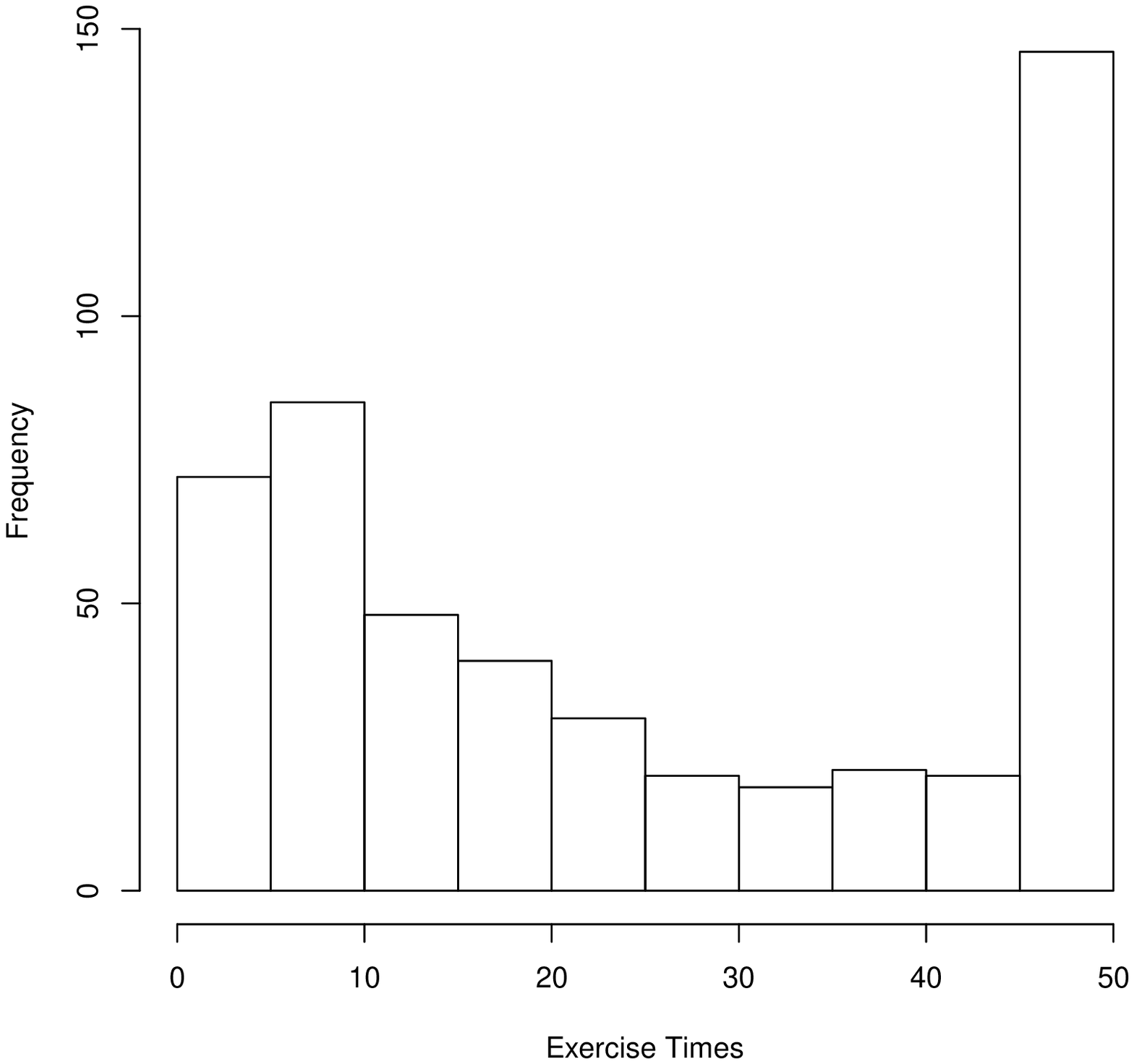}
  \caption{Distribution of cumulated rewards and exercise times.}
  \label{bermuda_plot2}
\end{figure}

Figure \ref{bermuda_plot2} contains the histograms for the cumulated
rewards and exercise times. Let us emphasise the usefulness of such
scenario generation. Given an approximately optimal policy and
backtesting, one can perform statistical analysis on the backtested
values to obtain practical insights such as for risk analysis
purposes.

\section{Example: Swing Option}

Let us now consider the swing option which is a financial contract
popular in the energy business.  In the simplest form, it gives the
owner the right to obtain a certain commodity (such as gas or
electricity) at a pre-specified price and volume at a number of
exercise times which can be freely chosen by the contract owner. Let
us consider a specific case of such contract, referred to as a
{unit-time refraction period} swing option.  In this contract, there
is a limit to exercise only one right at any time. Given the
discounted commodity price $(S_{t})_{t=0}^{T}$, the so-called fair
price of a swing option with $N$ rights is given by the supremum
$$
\sup_{0\leq\tau_1<\dots<\tau_N\leq T}\mathbb{E}\Big{[}\sum^N_{n=1}(S_{\tau_n}-Ke^{-\rho {\tau_n}})^+\Big{]}
$$ 
over all stopping times $\tau_1,\dots,\tau_N$ with values in
$\{0, \dots, T\}$.  In order to represent this control problem as a
switching system, we use the position set $\mathbf{P}=\{1,\dots,N+1\}$
to describe the number of exercise rights remaining.  That is
$p \in \mathbf{P}$ stands for the situation when there are $p-1$
rights remaining to be exercised.  The action set $\mathbf{A}=\{1,2\}$
represents the choice between exercising ($a=1$) or not exercising
($a=2$). The control matrices $(\alpha_{p,p'}^{a})$ are given for
exercise action $a=1$
$$
\alpha_{p, p'}^{1}=\left\{ \begin{array}{ll} 1 & \text{if  $p'=1 \vee (p-1)$} \\
0 & \text{else,}
\end{array}\right.
$$
and for not-exercise action $a=2$ as
$$
\alpha_{p, p'}^{2}=\left\{ \begin{array}{ll} 1 & \text{if  $p'=p$} \\
0 & \text{else}
\end{array}\right.
$$
for all $p, p' \in
\mathbf{P}$. In the case of the swing option, the transition between
$p$ and
$p'$ occurs deterministically, since once the controller decides to
exercise the right, the number of rights remaining is diminished by
one.  The deterministic control of the discrete component is easier to
describe in therm of the matrix $(\alpha(p,a))_{p \in \mathbf{P}, a \in
  \mathbf{A}}$ where $p'=\alpha(p, a) \in
\mathbf{P}$ stands for the discrete component which is reached from $p \in
\mathbf{P}$ by the action $a \in
\mathbf{A}$.  For the case of the swing option this matrix is
\begin{equation*}
(\alpha(p,a))_{p \in \mathbf{P},a \in \mathbf{A}} =
\left[ \begin{array}{cc}
1 & 1 \\
1 & 2\\
2& 3 \\
\dots & \dots \\
N & N+1 \\
\end{array} \right]. \label{twodim}
\end{equation*}
Having modelled the discounted commodity price process as an
exponential mean-reverting process with a reversion parameter $\kappa
\in [0, 1[$, long run mean $ \mu>0$ and volatility $\sigma>0$, we
obtain the logarithm of the discounted price process as
\begin{equation*}
\tilde Z_{t+1} = (1-\kappa) (\tilde Z_{t}-\mu)+\mu + \sigma \epsilon_{t+1}, \quad \tilde Z_{0}=\ln(S_{0}). \label{ou}
\end{equation*}
A further transformation of the state space is required before linear
state dynamics can be achieved. If we introduce an augmentation with 1
via
$$
Z_{t}=\left[\begin{array}{c} 1 \\ \tilde Z_{t} \end{array}\right], \qquad  t=0, \dots, T.
$$
then it becomes possible to represent the evolution as the linear
state dynamics
$$
Z_{t+1}=W_{t+1}Z_{t}, \qquad t=0, \dots, T-1
$$
with independent and identically distributed matrix-valued random
variables $(W_{t})_{t=1}^{T}$ given by
$$
W_{t+1}= \left[ \begin{array}{cc} 1 & 0 \\  \kappa \mu + \sigma \epsilon_{t+1} &
(1 - \kappa) \end{array}\right], \quad t=0,...,T-1.
$$
The reward and scrap values are given by
\begin{equation} \label{eq:reward2}
r_t(p,(z^{(1)}, z^{(2)}),a)= (e^{z^{(2)}}-Ke^{-\rho t})^{+} \big{(} p-\alpha(p,a) \big{)}
\end{equation} 
for $t=0,\dots,T-1$ and
\begin{equation} \label{eq:scrap2}
r_T(p,(z^{(1)}, z^{(2)}))=(e^{z^{(2)}}-Ke^{-\rho T})^{+} \big{(} p-\alpha(p,1) \big{)}
\end{equation}respectively for all $p \in \mathbf{P}$ and $a \in \mathbf{A}$.

\subsection{Code Example}

In this example, consider a swing option with $5$ rights exercisable
on $101$ time points.  As before, we begin by performing the value
function approximation. On our machine, the following code takes
around $0.4$ cpu seconds or around $0.15$ real world seconds to run.

\begin{lstlisting}[caption = {Value function approximation}]
library(rcss)
## Parameters
rho <- 0
kappa <- 0.9
mu <- 0
sigma <- 0.5
K <- 0 
n_dec <- 101  ## number of time epochs
N <- 5  ## number of rights
n_pos <- N + 1 ## number of positions
grid <- cbind(rep(1, 101), seq(-2, 2, length = 101))  ## Grid
## Control matrix
control <- cbind(c(1, 1:N), 1:(N + 1))
## Reward subgradient representation
reward <- array(0, dim = c(101, 2, 2, nrow(control), n_dec - 1))
slope <- exp(grid[, 2])
for (tt in 1:(n_dec - 1)) {
    discount <- exp(-rho * (tt - 1))
    for (pp in 2:n_pos) {
        intercept <- (exp(grid[,2]) - K * discount) - slope * grid[, 2]
        reward[, 1, 1, pp, tt] <- intercept
        reward[, 2, 1, pp, tt] <- slope
    }
}
## Scrap subgradient representation
scrap <- array(0, dim = c(101, 2, nrow(control)))
discount <- exp(-rho * (n_dec - 1))
for (pp in 2:n_pos) {
    intercept <- (exp(grid[,2]) - K * discount) - slope * grid[, 2]
    scrap[, 1, pp] <- intercept
    scrap[, 2, pp] <- slope
}
## Disturbance sampling
weight <- rep(1/1000, 1000)
disturb <- array(0, dim = c(2, 2, 1000))
disturb[1, 1,] <- 1
disturb[2, 2,] <- 1 - kappa
CondExpected <- function(a, b){
    return(1/sqrt(2 * pi) * (exp(-a^2/2)- exp(-b^2/2)))
}
part <- qnorm(seq(0, 1, length = 1000 + 1))
for (i in 1:1000) {
    disturb[2,1,i] <- kappa * mu + sigma * (CondExpected(part[i], part[i+1]) / (pnorm(part[i+1]) - pnorm(part[i])))
}
## Bellman recursion
r_index <- matrix(c(2, 1), ncol = 2)
bellman <- FastBellman(grid, reward, scrap, control, disturb, weight, r_index)
\end{lstlisting}

After obtaining these function approximations, the following code
computes the $99\%$ confidence intervals for the value of a swing
option with $5$ remaining rights. The code below takes approximately
$20$ cpu seconds or $10$ real world seconds to run.

\begin{lstlisting}[caption = {Lower and upper bounds}]
## Exact reward function
RewardFunc <- function(state, time) {
    output <- array(0, dim = c(nrow(state), 2, nrow(control)))
    discount <- exp(-rho * (time - 1))
    for (i in 2:nrow(control)) {
        output[, 1, i] <- pmax(exp(state[, 2]) - K * discount, 0)
    }
    return(output)
}
## Exact scrap function
ScrapFunc <- function(state) {
    output <- array(0, dim = c(nrow(state), nrow(control)))
    discount <- exp(-rho * (n_dec - 1))
    for (i in 2:nrow(control)) {
        output[, i] <- pmax(exp(state[, 2]) - K * discount, 0)
    }
    return(output)
}
## Generate paths
set.seed(12345)
n_path <- 500
path_disturb <- array(0, dim = c(2, 2, n_path, n_dec - 1))
path_disturb[1, 1,,] <- 1
path_disturb[2, 2,,] <- 1 - kappa
rand1 <- rnorm(n_path * (n_dec - 1) / 2)
rand1 <- as.vector(rbind(rand1, -rand1))
path_disturb[2, 1,,] <- kappa * mu + sigma * rand1
start <- c(1, 0)
path <- PathDisturb(start, path_disturb)
policy <- FastPathPolicy(path, grid, control, RewardFunc, bellman$expected)
## Set subsimulation disturbances
n_subsim <- 500
subsim <- array(0, dim = c(2, 2, n_subsim, n_path, n_dec - 1))
subsim[1, 1,,,] <- 1
subsim[2, 2,,,] <- 1 - kappa
rand2 <- rnorm(n_subsim * n_path * (n_dec - 1) / 2)
rand2 <- as.vector(rbind(rand2, -rand2))
subsim[2, 1,,,] <- kappa * mu + sigma * rand2
subsim_weight <- rep(1 / n_subsim, n_subsim)
## Primal-dual
mart <- FastAddDual(path, subsim, subsim_weight, grid, bellman$value, ScrapFunc)
bounds <- AddDualBounds(path, control, RewardFunc, ScrapFunc, mart, policy)
\end{lstlisting}

\begin{lstlisting}[caption = {99\% confidence interval}]
> print(GetBounds(bounds, 0.01, 6))
[1] 13.42159 13.44162
\end{lstlisting}

\section{Conclusion}
\label{sec_end}

This paper gives a demonstration of the R package \emph{rcss} in
solving optimal switching problems. The problem setting discussed in
this paper is broad and can be used to model a wide range of problems.
Using nearest neighbour algorithms, the package \emph{rcss} is able
to solve some real world problems in an accurate and quick manner.

\bibliography{main}
\bibliographystyle{amsplain}

\end{document}